\documentclass[sigconf]{acmart}
\usepackage{CJKutf8}
\usepackage{amsmath, amssymb, amsfonts}
\AtBeginDocument{%
  }

\usepackage{array} 
\usepackage{geometry} 
\usepackage{longtable}
\usepackage{tabularx}
\usepackage{graphicx}
\usepackage{multirow}
\usepackage{booktabs}
\usepackage{tabulary}
\usepackage{makecell}
\usepackage{hyperref}
\usepackage{amssymb}
\usepackage{tikz}
\usepackage{enumitem} 

\newcommand{\hl}[1]{#1}
\definecolor{bad}{HTML}{AD0000}
\definecolor{good}{HTML}{107260}

\copyrightyear{2025}
\acmYear{2025}
\setcopyright{acmlicensed}\acmConference[CHI '25]{CHI Conference on Human Factors in Computing Systems}{April 26-May 1, 2025}{Yokohama, Japan} 
\acmBooktitle{CHI Conference on Human Factors in Computing Systems (CHI '25), April 26-May 1, 2025, Yokohama, Japan} 
\acmDOI{10.1145/3706598.3715271}
\acmISBN{979-8-4007-1394-1/25/04}
  
\begin{document}
\begin{CJK*}{UTF8}{gbsn}

\title{Walk in Their Shoes to Navigate Your Own Path: Learning About Procrastination Through A Serious Game}



\author{Runhua Zhang}
\authornote{Runhua Zhang is now a Ph.D. student at Hong Kong University of Science and Technology; this work was done during her post-graduate studies at Tongji University.}
\email{runhua.zhang@connect.ust.hk}
\affiliation{%
  \institution{Tongji University}
  \city{Shanghai}
  \country{China}}
\affiliation{%
  \institution{Hong Kong University of Science and Technology}
  \city{Hong Kong SAR}
  \country{China}}

\author{Jiaqi Gan}
\email{jiaqigan.kiki@gmail.com}
\affiliation{%
  \institution{Independent Researcher}
  \state{California}
  \country{United States}}

\author{Shangyuan Gao}
\email{ciel_@tongji.edu.cn}
\affiliation{%
 \institution{Tongji University}
 \city{Shanghai}
 \country{China}}

\author{Siyi Chen}
\email{2152689@tongji.edu.cn}
\affiliation{%
 \institution{Tongji University}
 \city{Shanghai}
 \country{China}}

\author{Xinyu Wu}
\email{wuxinyu2003@tongji.edu.cn}
\affiliation{%
 \institution{Tongji University}
 \city{Shanghai}
 \country{China}}

\author{Dong Chen}
\email{eastchen@tongji.edu.cn}
\affiliation{%
 \institution{Tongji University}
 \city{Shanghai}
 \country{China}}

\author{Yulin Tian}
\email{yulin_t@tongji.edu.cn}
\affiliation{%
 \institution{Tongji University}
 \city{Shanghai}
 \country{China}}

\author{Qi Wang}
\email{qiwangdesign@tongji.edu.cn} 
\authornote{Pengcheng An and Qi Wang are the corresponding authors}
\affiliation{%
 \institution{Tongji University}
 \city{Shanghai}
 \country{China}}

\author{Pengcheng An}
\email{anpc@sustech.edu.cn}
\affiliation{%
 \institution{Southern University of Science and Technology}
 \city{Shenzhen}
 \country{China}}
\authornotemark[2]

\renewcommand{\shortauthors}{Zhang et al.}

\begin{abstract}
Procrastination, the voluntary delay of tasks despite potential negative consequences, has prompted numerous time and task management interventions in the HCI community. While these interventions have shown promise in addressing specific behaviors, psychological theories suggest that learning about procrastination itself may help individuals develop their own coping strategies and build mental resilience. However, little research has explored how to support this learning process through HCI approaches. We present \textit{ProcrastiMate}, a text adventure game where players learn about procrastination's causes and experiment with coping strategies by guiding in-game characters in managing relatable scenarios. Our field study with 27 participants revealed that \textit{ProcrastiMate} facilitated learning and self-reflection while maintaining psychological distance, motivating players to integrate newly acquired knowledge in daily life. This paper contributes empirical insights on leveraging serious games to facilitate learning about procrastination and offers design implications for addressing psychological challenges through HCI approaches.
\end{abstract}

\ccsdesc[500]{Human-centered computing~Empirical studies in HCI}

\keywords{Procrastination, Serious Games, Learning, Reflection}

\begin{teaserfigure}
  \includegraphics[width=\textwidth]{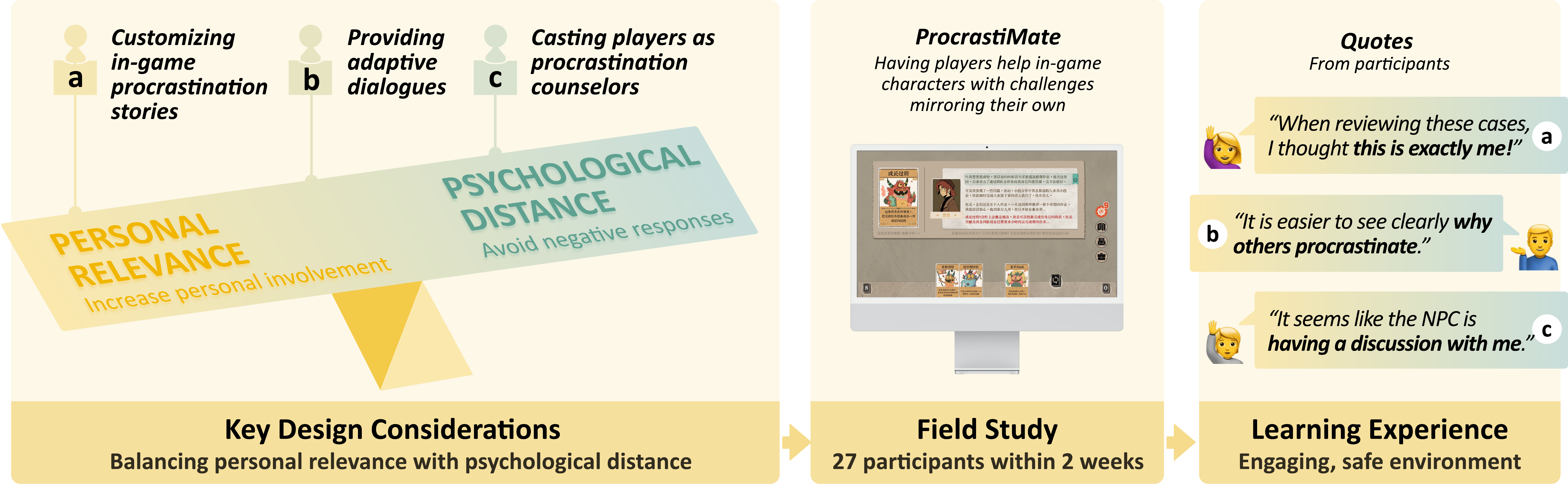}
  \caption{In this study, we designed and evaluated ProcrastiMate, a serious game to help players learn about procrastination. The key design challenge was balancing \textcolor[HTML]{F1B600}{personal relevance} with \textcolor[HTML]{5EA191}{psychological distance}—allowing players to \textcolor[HTML]{F1B600}{deeply engage} \textcolor[HTML]{5EA191}{without feeling judged} (as shown on both ends of the scale). We addressed this by: a) customizing in-game procrastination stories, b) providing adaptive dialogues, and c) casting players as procrastination counselors. By \textcolor[HTML]{5EA191}{having players help in-game characters} with \textcolor[HTML]{F1B600}{challenges mirroring their own}, ProcrastiMate created a safe and engaging environment for effective learning and reflection.}
  \label{teaser}
\end{teaserfigure}


\maketitle

\section{Introduction}
Procrastination, putting off tasks needed to do or sometimes not doing them at all, is often considered as a manifestation of poor time management skills \cite{senecal1995self, steel2007nature}. It is estimated that approximately 20\% of the adult population and between 80\% to 95\% of college students would procrastinate to some extent \cite{turner2023systematic}. In response to its prevalence among the colleges, the Human-Computer Interaction (HCI) community has developed various technological interventions aimed at reducing specific procrastination behaviors, such as scheduling tools \cite{higashi2023personalized}, to-do list applications \cite{pereira2021struggling,valladares2016designing,kirchner2024outplay}, and project management applications \cite{arakawa2023catalyst}. While these interventions have shown promising results in improving productivity and task completion, they primarily address the issue from the perspectives of time management and task organization.

However, psychological theories suggest that procrastination also stems from deeper emotional and psychological factors \cite{beswick1988psychological}, such as fear of failure \cite{schraw2007doing,flett2012procrastination,beck2000correlates}, perfectionism \cite{rice2012perfectionism}, and self-protective mechanisms \cite{sirois2013procrastination}. Addressing these psychological and emotional aspects requires helping individuals understand the underlying causes of their procrastination, which can be a crucial step toward effective managements \cite{solomon1984academic,beck2000correlates,glick2014preliminary}. Traditional educational approaches, including workshops and seminars for college students, have shown promise in facilitating this learning process \cite{glick2015investigation,rozental2014group}. By gaining insights into procrastination, individuals can recognize the causes of their behavior and adopt coping strategies tailored to their situations, thereby gaining more autonomy \cite{steel2016academic}. 

Despite these insights, there remains a gap in how to effectively support individuals in learning about procrastination through HCI approaches, which offer more interactive and personalized solutions compared to traditional educational methods. To address this research gap, our study utilized serious game as a vehicle, leveraging their ability to encourage active engagement and \hl{make complex topics more approachable} in a controlled, simulated environment \cite{ganesh2023tailoring,bauer2023improving,nilsen2020playing}. Specifically, we designed and evaluated a text adventure game named \textit{ProcrastiMate}. The game aims to facilitate players in learning about procrastination's causes and experimenting with coping strategies, while providing an engaging and safe learning experience. 

\hl{\textit{ProcrastiMate} was informed by procrastination theories as well as the insights from our formative study.} Set in a \hl{fictional} university, players assume the role of a counselor tasked with transforming the institution's traditional punitive approaches to managing procrastination into a more supportive environment. The core gameplay involves helping students \hl{(non-player characters, NPCs)} tackle their procrastination challenges by identifying underlying causes and suggesting appropriate coping strategies. \textit{ProcrastiMate} encourages players to explore and experiment with new concepts by helping in-game characters, with the aim of fostering psychological distance. Meanwhile, by mirroring the players' own procrastination challenges through those of the NPCs, the game seeks to offer highly relatable narratives that facilitate personal relevance and self-reflection.

We then evaluated \textit{ProcrastiMate} to understand 1) whether and how \textit{ProcrastiMate} supports players in learning about procrastination, and 2) how specific design considerations facilitate the learning process and experiences. A field study within a two-week period was conducted with 27 participants, incorporating pre- and post-game semi-structured interviews along with quantitative measures as auxiliary references. Our findings revealed that \textit{ProcrastiMate} \hl{offered an enjoyable way for players to develop a deeper understanding of procrastination while fostering self-reflection on their own behavior. Moreover, our findings highlighted how the effective learning process was facilitated by creating an appropriate psychological distance with players, making the game relatable yet non-threatening, allowing them to engage without fear of criticism}. 

\hl{The contribution of this paper is threefold. First, our formative study revealed the importance of acknowledging the negative emotions tied to procrastination and highlighted the need \textbf{to balance psychological distance with personal relevance} in educational approaches for supportive learning. Second, informed by these insights, we \textbf{designed and evaluated \textit{ProcrastiMate}}, a text adventure game that offers a psychologically safe space for exploring the causes of procrastination and practicing coping strategies. Unlike behavior-oriented interventions in HCI, \textit{ProcrastiMate} prioritized engaging learning and self-reflection, while also addressing the limited accessibility and adaptability of traditional educational methods. Finally, we discussed \textbf{design implications}: 1) leveraging playful, educational approaches to break the procrastination cycle, 2) designing effective learning experience with serious games to address sensitive psychological challenges beyond procrastination, and 3) addressing potential challenges in educational interventions.}

\section{Related Work}
\hl{In this section, we outline the motivation behind our approach. We begin by understanding procrastination, emphasizing the potential of educational approaches for its intervention while acknowledging the limitations of traditional methods. Next, we review previous HCI interventions, and identified a gap in utilizing HCI for procrastination education that fosters learning and reflection. Lastly, we argue that serious games offer a promising direction for addressing this gap.}

\subsection{Understanding Procrastination: From Behavioral Management to Psychological Insights}
\label{2.1}
Procrastination, despite its varied definitions, can be described as \textit{the voluntary delay of an intended course of action despite expecting to be worse off for the delay} \cite{steel2007nature}. It is generally regarded as a failure of self-regulation \cite{rebetez2016procrastination}, often attributed to lack of self-control \cite{schouwenburg2001study} or poor time management skills \cite{wolters2017examining}. Traditional interventions aim to improve self-regulation by bridging the gap between intentions and actions \cite{van2005bridging}. These typically include time management strategies like scheduling and prioritization \cite{lay1993trait, claessens2007review, gieselmann2016treating, hafner2014avoiding}, habit formation through implementation intentions (‘if-then’ plans) \cite{owens2008overcoming, wieber2010overcoming, ajzen2009intentions}, and removing distractions \cite{van2000procrastination, mulry1994psychological}. Previous interventions in the HCI community have similarly focused on these approaches (detailed in \autoref{hci_intervention}). 

However, another substantial body of research has focused on understanding psychological causes and mechanisms of procrastination \cite{solomon1984academic, beck2000correlates, glick2014preliminary}, revealing that procrastination is not solely a result of poor time management or lack of willpower \cite{van2018overcoming}. \hl{One widely accepted framework for understanding procrastination, proposed by Piers Steel \cite{steel2007nature}, identifies four key causes: 1) \textbf{low self-efficacy}, where individuals lack confidence in their ability to complete tasks, often perceiving them as overly difficult or feeling inadequate despite effort; 2) \textbf{low perceived task value}, including disliking the task, finding it meaningless, or perceiving it as unchallenging or uninteresting; 3) \textbf{high impulsiveness}, marked by difficulty resisting distractions, such as entertainment or mobile phone use; and 4) \textbf{distant delay}, related to the temporal distance of rewards and punishments, leading to underestimating deadlines, procrastinating on non-urgent tasks, and struggling with long-term planning.}

\hl{In this context, helping individuals understand the causes and mechanisms of procrastination is considered an alternative intervention}. Proponents, such as Steel and Klingsieck \cite{steel2016academic}, argue that understanding procrastination is the key to overcoming it. Specifically, by enhancing the understanding, individuals not only gain awareness of the reasons behind their procrastination \cite{rozental2014understanding} but also improve their emotional regulation \cite{sirois2015less, pychyl2016procrastination}. This improved regulation would further help address the psychological discomfort associated with task avoidance, fostering resilience against procrastination tendencies \cite{glick2014preliminary}, and has the potential to enable more effective and long-term responses  \cite{grunschel2018ll}.

\hl{Despite the benefits of educational approaches, traditional methods in education and psychoeducation—such as seminars and lectures for college students—are often less accessible and tend to follow a one-size-fits-all format, limiting their adaptability to individual needs. In light of these limitations, HCI approaches offer promising alternatives by leveraging interactive and personalized mediums, which motivates our current work.} 

\subsection{Procrastination Interventions in HCI}
\label{hci_intervention}
\hl{Previous HCI studies on procrastination interventions predominantly relied on psychological insights to improve users' motivation for action or enhance their self-regulation to manage procrastination. These interventions often involved motivational strategies and tools such as project management, to-do lists, or reminders}. For example, Valladares et al. incorporated gamification elements, such as bet-placing and financial rewards, to motivate users to track tasks \cite{valladares2016designing}. Wu et al. developed a to-do list that leveraging social networks to increase engagement by incorporating peer-driven excitement \cite{wu2021dillydally}. Similarly, Higashi et al. introduced personalized agents on Slack\footnote{\url{https://en.wikipedia.org/wiki/Slack_(software)}} to encourage adherence to planned schedules \cite{higashi2023personalized}. A recent study by Kirchner-Krath et al. compared gamified and non-gamified task management apps, finding that gamification can boost motivation and help overcome procrastination \cite{kirchner2024outplay}. Earlier studies also explored using SMS reminders to prompt task planning \cite{edwards2015examining}.

Recent advancements in chatbots and large language models (LLMs) offer new opportunities for providing personalized task management suggestions, expanding on the tools mentioned earlier. For instance, Pereira and Díaz developed a chatbot that provides project management suggestions to build self-regulatory skills and self-efficacy \cite{pereira2021struggling}. Bhattacharjee et al. explored how LLMs can provide personalized, structured action steps and deadline-driven instructions to address academic procrastination \cite{bhattacharjee2024understanding}.

In addition to time or task management, some interventions aimed to reduce distractions, further helping users stay on task. The CatAlyst extension, for instance, used LLMs to generate content for writing and editing tasks, aiding users in resuming work and reducing procrastination \cite{arakawa2023catalyst}. Similarly, Aiki employed a ‘redirection of activity’ strategy, guiding users to educational websites before granting access to time-wasting sites \cite{inie2021aiki}.

\hl{While these interventions have demonstrated their effectiveness in modifying behavior, few have adopted an educational approach that fosters learning and reflection on procrastination. However, as highlighted in \autoref{2.1}, educational methods hold promising potential for increasing self-awareness, reducing negative emotions, and building resilience. Motivated by this gap, we aimed at exploring HCI approaches, such as serious games that are often used for awareness-building and reflection facilitation, to help individuals better understand procrastination itself.}

\subsection{Serious Games for Learning and Reflection} 
\label{2.3}
\hl{The HCI community has developed various approaches to facilitate learning and awareness-building, from e-learning platforms or applications \cite{holly2024gamedevdojo}, to personal informatics \cite{liao2022nudge} and chatbots \cite{hedderich2024piece}. Among these, serious games—designed for purposes beyond entertainment \cite{michael2005serious}—have demonstrated particular effectiveness across diverse domains, such as emotional regulation \cite{fan2018emostory}, political discussion \cite{rajadesingan2023guessync}, health behavior \cite{pasumarthy2024go, orji2017towards}, or disability awareness education \cite{jin2023divrsity, gerling2014effects}. } 

\begin{figure*}[t]
  \centering
  \includegraphics[width=\linewidth]{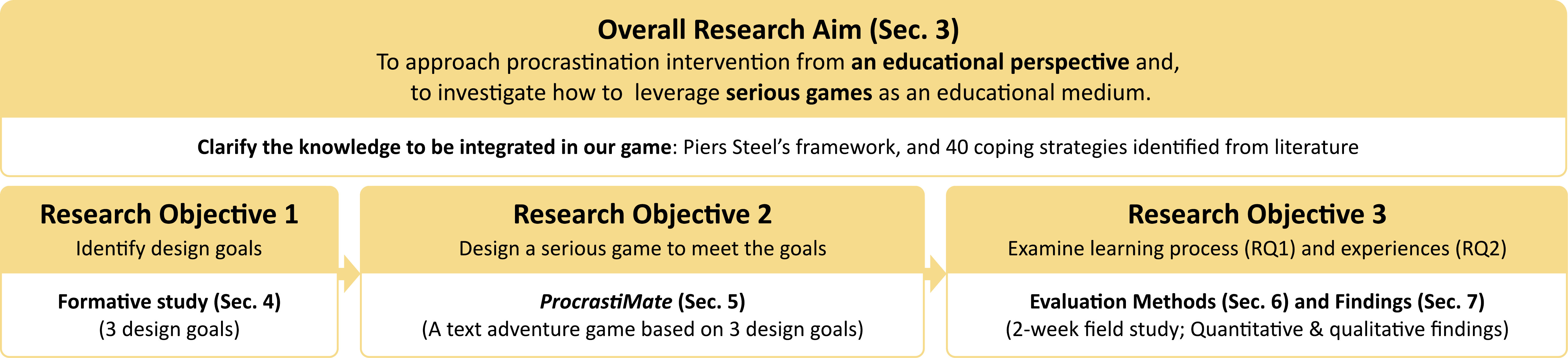}
  \caption{The overall research aim of our study and the corresponding research pipeline}
  \label{overall}
\end{figure*}

\hl{In our study, we aimed to leverage serious games to approach psychologically complex topics like procrastination due to their three main advantages. First, serious games can create playful learning environments that make challenging or stressful subjects more approachable, as demonstrated in contexts like children's emotion management \cite{fan2018emostory} and cancer treatment discussions \cite{gerling2011designing}. This is particularly valuable for addressing procrastination, where resistance to confronting the issue may create barriers to learning. Second, serious games feature low-stakes environments for experimentation through the elements like role playing, dialogues, and in-game storytelling \cite{book}, allowing players to safely explore new perspectives without real-world consequences \cite{heng2021rewind, ashby2023personalized}. Finally, serious games can simulate real-world decision-making, linking abstract concepts to everyday experiences, as exemplified by ‘What.Hack’ in educating phishing identification \cite{wen2017hack}. This can be crucial for procrastination, where abstract psychological concepts learned are expected to be connected with everyday behavior.}

\hl{Despite these strengths and their effectiveness in various psychological and behavioral contexts, using serious games as a medium for educational procrastination interventions remains largely unexplored, motivating our investigation in this direction.}

\section{Research Aim}
\hl{As outlined above, our research aims to approach procrastination intervention from an educational perspective, leveraging serious games as a medium. This aim is supported by three objectives: 1) identify key design challenges and goals for creating a serious game that fosters learning and reflection on procrastination, 2) design and implement a game based on these goals, and 3) examine the learning processes, outcomes, and player experiences, thereby contributing to the HCI community's understanding of this issue.}

\hl{To achieve these objectives, \textbf{we first clarify the theoretical foundation (i.e., key knowledge) to be integrated into the game}. We grounded our game design in the widely recognized framework proposed by Piers Steel \cite{steel2007nature}, which, as discussed in \autoref{2.1}, identifies four primary causes of procrastination: 1) low self-efficacy, 2) low perceived task value, 3) high impulsiveness, and 4) distant delay. In alignment with Steel's framework, we also identified ten coping strategies for each of the four causes through an extensive literature review on procrastination management\footnote{See our supplementary for the references for each coping strategy}. Both the framework and 40 coping strategies guided our subsequent game design decisions.} \hl{\textbf{We then adopted a research-through-design approach}. As shown in \autoref{overall}, we began with a formative study (\autoref{formative study section}) using initial gameplay elements to identify design goals for a full version game. We then designed and implemented our serious game, \textit{ProcrastiMate} (\autoref{procrastimate section}). Finally, we detailed our evaluation methods (\autoref{evaluation section}) and presented both quantitative and qualitative findings (\autoref{findings section}).}

\begin{figure*}[t]
  \centering
  \includegraphics[width=\linewidth]{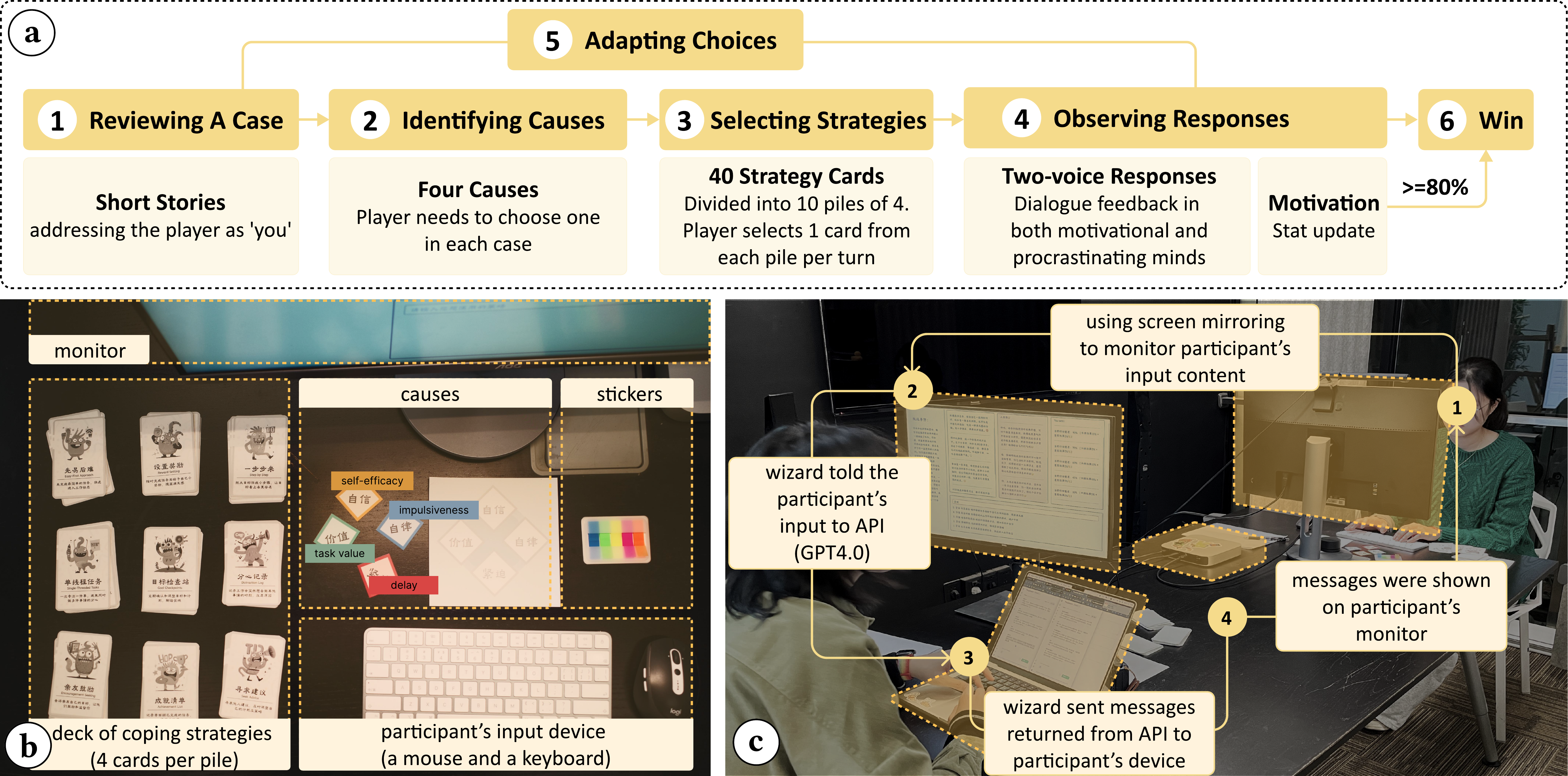}
  \caption{The initial gameplay and the formative study settings: a) The gameplay designed for formative study with initial design elements, b) The physical components prepared for participants; c)The formative study setup.}
  \label{formative study}
\end{figure*}

\section{Formative Study}
\label{formative study section}
\hl{To address our first research objective, we conducted a formative study focusing on college students as our target users. This demographic often struggles with managing heightened autonomy while navigating complex trade-offs between academic, social, and personal responsibilities—factors that exacerbate procrastination’s effects on both academic performance and psychological well-being \cite{zhu2023positive}. In this section, we present our initial design elements, study methods, key findings from the college student participants' and experts' feedback, and three identified design goals.}

\subsection{Initial Design}
Inspired by the negotiation mechanics in \textit{Griftlands}\footnote{\url{https://en.wikipedia.org/wiki/Griftlands}} and \textit{Disco Elysium}\footnote{\url{https://en.wikipedia.org/wiki/Disco_Elysium}}, which allows players to achieve objectives through dialogue strategies, our design emphasized strategic communication for problem-solving. Specifically, the initial design aimed to immerse players in procrastination scenarios, guide them in identifying the underlying causes, and use appropriate coping strategies to address them. To achieve this, we employed two key design elements: 

\textbf{1) Procrastination scenarios presented through short stories}. We used short stories to introduce players to various procrastination scenarios. Written in the second-person pronoun, these stories engage the player directly with phrases such as “\textit{you face a procrastination problem...}” to foster personal involvement. The narratives were crafted based on the four causes of procrastination identified by Steel\cite{steel2007nature}, and set in typical university settings—like procrastinating on assignments for long-term projects or elective courses—to enhance personal relevance.

\textbf{2) Card decks:} \hl{Leveraging the coping strategies identified, we introduced a deck of 40 cards with each card representing a title, an explanation, and its utility. For example, Card No. 1 was titled “\textit{Step by Step}”, with an explanation “\textit{Break large tasks into smaller segments to increase confidence in completing each smaller task}.” Details on all 40 cards are provided in \autoref{appendix_card},} where cards No. 1-10 are involved to improve self-efficacy, No. 11-20 aim to enhance perceived task value, No. 21-30 help control impulsiveness, and No. 31-40 address distant deadlines.

Based on the design elements, as shown in \autoref{formative study}-a, a preliminary gameplay sequence was developed to promise a game experience. The gameplay unfolds as follows: 1) Players begin by reviewing short stories, each with an initial ‘Motivation’ stat of 20\%. 2) They analyze the case to identify underlying causes before selecting cards. 3) In each turn, players choose one coping strategy card from four options, aiming to address the identified causes and increase the Motivation stat. 4) The game responds with contextual dialogues from two narrative voices: a motivational mind encouraging task completion, and a procrastinating mind voicing excuses. Besides, the Motivation stat increases by 5\%, 10\%, or 15\%, depending on the cause (+5\% if correct) and card choice (+10\% if correct). 6) Players adapt their strategy based on the feedback, selecting subsequent cards to refine their approach. 6) The game continues until the Motivation stat exceeds 80\%, at which point the player wins the case.

\subsection{Methods}
\hl{Employing the initial gameplay, we developed a paper prototype for a Wizard-of-Oz study with 5 potential players and 2 experts, to gather feedback on our design. This section details participant recruitment, setup, procedures and data collection, and data analysis. The study received the ethical approval from the first author's institution.}

\subsubsection{Participants}
We recruited participants from public online chat groups at a university in China, targeting individuals who self-identify as procrastinators interested in learning more about the topic. Our study involved seven participants in total: five potential players (G1 - G5; three females, two males, with other gender options provided) and two game design experts. One expert was a UI/UX designer (E1, female) from a renowned game company in China, and the other was a game designer (E2, male) from a university game research lab. All participants were in their 20s, reported no mental health diagnoses, and had not used psychotropic medications in the past three months. Each participant received a café voucher worth 40 CNY upon completing the study. 

\subsubsection{Setup} A paper prototype based on the initial design was developed, to be used in a controlled laboratory environment with a Wizard-of-Oz method. The prototype consisted of: \textbf{1) Physical Components} (\autoref{formative study}-b): A deck of 40 coping strategy cards, printed to resemble standard playing cards, and printed 4 cause cards based on Steel's framework. \textbf{2) Digital Interface}: An interactive webpage displayed the procrastination stories and facilitated gameplay. It featured updates on two narrative voices (generated by OpenAI's GPT-4) and the Motivation stat, and provided an input space for players to type identified causes and select strategies. \textbf{3) Wizard Interface} (\autoref{formative study}-c): A separate monitor synchronized with the player's screen allowed the wizard to observe the content displayed to participants and send responses, making participants believe that they were interacting directly with the system.

\subsubsection{Procedures and Data Collection}
We began by explaining the purpose and procedures of the study, obtaining consent from participants for voice and screen recording. Participants were instructed to think aloud during gameplay, after which we provided detailed gameplay instructions. They then engaged with four procrastination cases using the paper prototype, each designed to represent one of Steel's four causes of procrastination, without any time limit for the gaming experience. \hl{During gameplay, the wizard responded in real-time based on the participants' verbalized thoughts and screen activity.} Each gameplay session lasted between 15 and 30 minutes, excluding the time for instructions.

Following the gameplay session, a semi-structured interview was conducted to explore participants' attitudes towards the game, feedback on gameplay elements, and expectations for the full game version. \hl{Example interview questions included: “\textit{What are your thoughts on the procrastination stories presented? How do you feel about presenting coping strategies in card form? What is your opinion on the gameplay, specifically using cards as dialogue strategies to address in-game procrastination?}”. After the interviews, we presented the initial design and player's feedback to the two experts to elicit their professional suggestions. All interviews lasted between 30 and 40 minutes. The full interview outlines for general users and experts are available in our Supplementary material.}

\subsubsection{\hl{Data Analysis}} \hl{All interviews were transcribed verbatim to ensure accuracy. We took an open coding approach ~\cite{hancock2001introduction} to inductively explore participants' gaming experiences and expectations regarding our initial design. Specifically, the first two authors served as coders, independently familiarized themselves with the data and independently generated initial codes. Then, through iterative discussions, the coders compared and refined their codes to establish a shared codebook. Examples of the resulting codes include “\textit{logic of playing cards}”, “\textit{attitudes toward in-game stories}”, and “\textit{real-time feedback}”. Finally, two coders applied the affinity diagramming \cite{lucero2015using}, grouping similar codes into three overarching themes relevant to our research objective 1: “\textit{relatable in-game stories}”, “\textit{player role}”, and “\textit{role of dialogue feedback}”. These themes were explained in the following findings.} 

\subsection{Findings}
In the interviews, all participants expressed interest in a full version of the game, recognizing its potential to enhance motivation for exploring new concepts related to procrastination and finding personal solutions. They appreciated the transformation of causes and coping strategies into game elements, such as short stories and a deck of cards, noting that these elements fostered continued engagement with the gameplay. Importantly, our formative study identified a key challenge for the full game: \textit{balancing 1) highly personal relevance with 2) the need to maintain sufficient psychological distance for effective learning}. \hl{It revealed that when players perceived the game as directly addressing their own procrastination issues, they often resisted new information that contradicted their existing beliefs. This insight was linked to the following three findings:}

\subsubsection{Finding 1: Relatable Narratives Foster the Connection Between In-game and Personal Experiences}
This finding indicated that realistic, relatable procrastination scenarios significantly enhanced players' engagement with the game's core mechanics. The majority of participants (6 out of 7) expressed strong personal connections to the in-game situations, making comments like \textit{“this is me!}” or “\textit{isn’t this exactly like me?!}” This personal identification encouraged a more thoughtful approach to cause identification and strategy selection within the game. G4's comment exemplified this connection: “\textit{When I felt as though the story was describing my own situation, I suddenly felt like it was a chance to re-evaluate my own behavior. So, I ended up spending more time thinking about the right choices.}” Expert feedback further inspired us regarding the future development: “\textit{How about designing customized stories for the players of this game? It may have the chance to provoke their deeper reflection on their own experience}.”

\subsubsection{Finding 2: Direct Procrastinator Role Assignment Triggers Defensive Learning Behaviors}
Our findings revealed a complex interplay between players' reliance on personal experiences and their receptiveness to new perspectives in the gameplay. We found that when the game directly implicated players in procrastination scenarios (e.g., “\textit{you are procrastinating}”), they consistently approached in-game situations by drawing upon their personal experiences. This approach, while fostering relatability, inadvertently limited players' exposure to potentially more effective strategies. For instance, G5 remarked, “\textit{I always set earlier deadlines for myself, and this worked well, so I would like to use this one}.” While setting earlier deadlines might increase a sense of urgency, it may not significantly impact one's self-efficacy, which is the primary cause of the case. This mismatch would bring frustration or disengagement to players. G5 further commented: “\textit{I thought I was solving my own problem in the game, but the useful strategies I used in my life did not work in this game. Sometimes I felt it was difficult for me to accept the feedback}.”

Moreover, we found that using direct address (e.g., “\textit{you}”) to encourage players to imagine themselves as procrastinators had unintended negative emotional consequences. This approach often led players to feel judged or experience negative emotional responses such as anxiety, guilt, or stress. G4 expressed this sentiment: “\textit{The feeling of judging myself on this issue makes me feel stressed}.” In discussing these observations, E1 invoked the proverb, “\textit{The bystander sees more of the game than the player},” suggesting that a more detached and objective experience might be beneficial. 

\subsubsection{Finding 3: Adaptive Dialogue Serves as A Scaffold for Facilitating Understanding of New Concepts}
We also found that real-time dialogue responses, particularly those offering critical perspectives, helped players understand the efficacy of various coping strategies within the game. Despite implementing a dual-voice response system, participants showed a marked preference for the more critical voice of the “procrastinating mind”. For example, G4 stated, “\textit{I paid more attention to the words from the procrastinating mind because they helped me think about why a strategy works, while the other does not.}” From an expert's viewpoint, E2 emphasized the importance of explanation in serious games: “\textit{If we want to involve players in a new perspective, it is important to help players understand the logic behind it...Maybe consider mixing the two minds, and what works and what does not work for a coping strategy can be presented at the same time}.” 

\begin{figure*}[t]
  \centering
  \includegraphics[width=\linewidth]{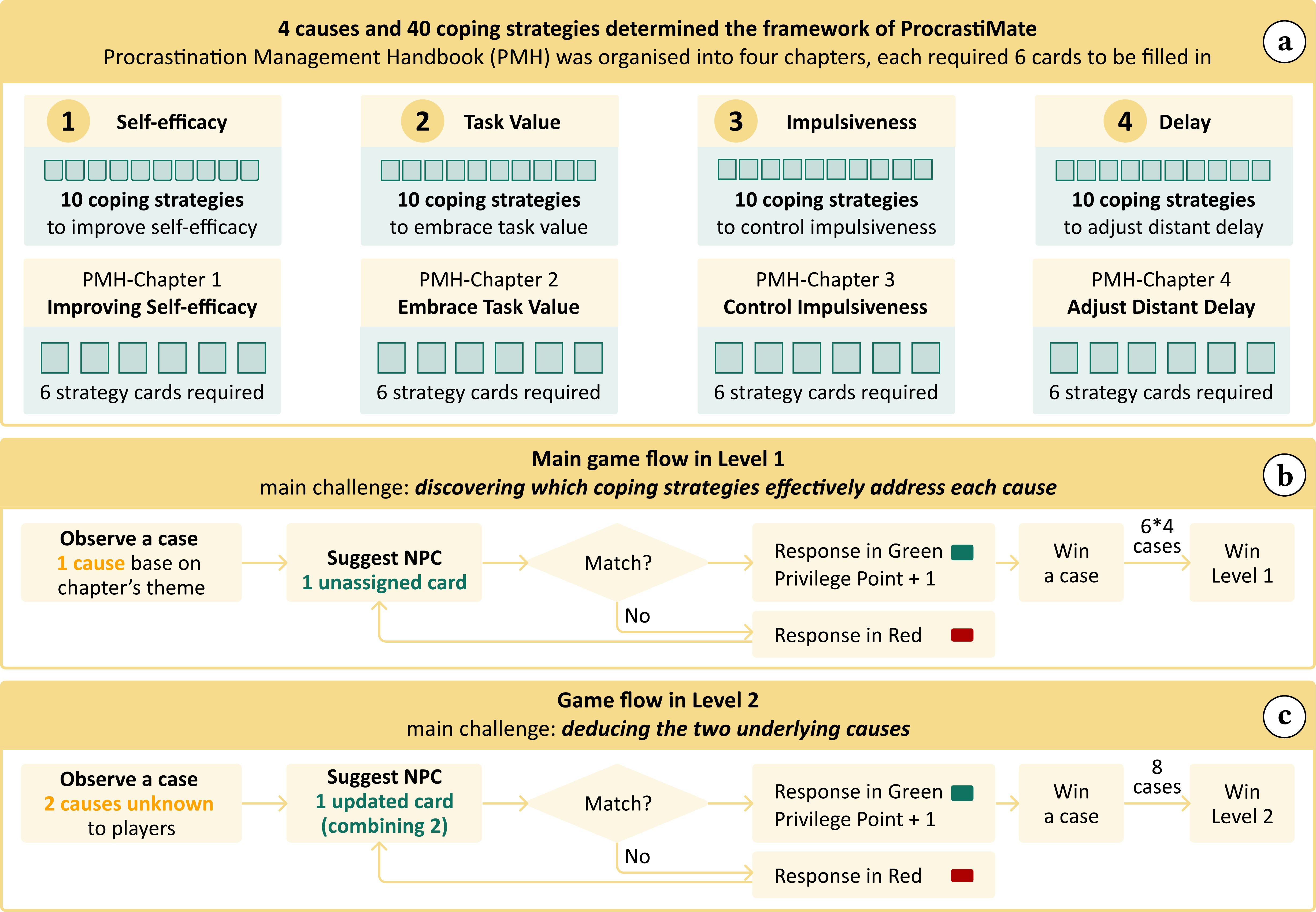}
  \caption{Overview of \textit{ProcrastiMate}. a) The game structure is based on four causes of procrastination and 40 coping strategies. The correspondences between causes and strategies guides the gameplay and determines the win conditions. The causes are organised into four chapters, each requiring six strategy cards to be filled in. b) Core gameplay mechanics in Level 1, challenging players to identify effective coping strategies for each cause; c) Gameplay in Level 2 requires players to diagnose the underlying causes of eight procrastination scenarios.}
  \label{fig:gameplay}
\end{figure*}

\subsection{Summary of Design Goals}
Based on the findings, we summarized that they highlighted the importance to maintaining the sufficient psychological distance while keeping personal involvement. Consequently, we have formalized three design goals for our full game.

\begin{itemize}
     \item[\hl{\textbf{[DG1]}}] \hl{\textbf{To enhance personal relevance}: Our game aimed to incorporate narratives and scenarios that resonated with players' real-life experiences to foster engagement and connection with the in-game content.}
     \item[\hl{\textbf{[DG2]}}] \hl{\textbf{To encourage engagement while maintaining objective distance}: Our game aimed to prompt players to approach procrastination scenarios from a relatively detached, analytical perspective, encouraging the exploration of new strategies beyond personal experiences while helping to avoid stress responses.}
     \item[\hl{\textbf{[DG3]}}] \hl{\textbf{To facilitate smooth knowledge integration}: Our game aimed to implement adaptive feedback and explanations to help players understand and gradually accept new concepts about procrastination, bridging the gap between their existing knowledge and the game's educational framework.}
\end{itemize}

\begin{figure*}[t]
  \centering
  \includegraphics[width=\linewidth]{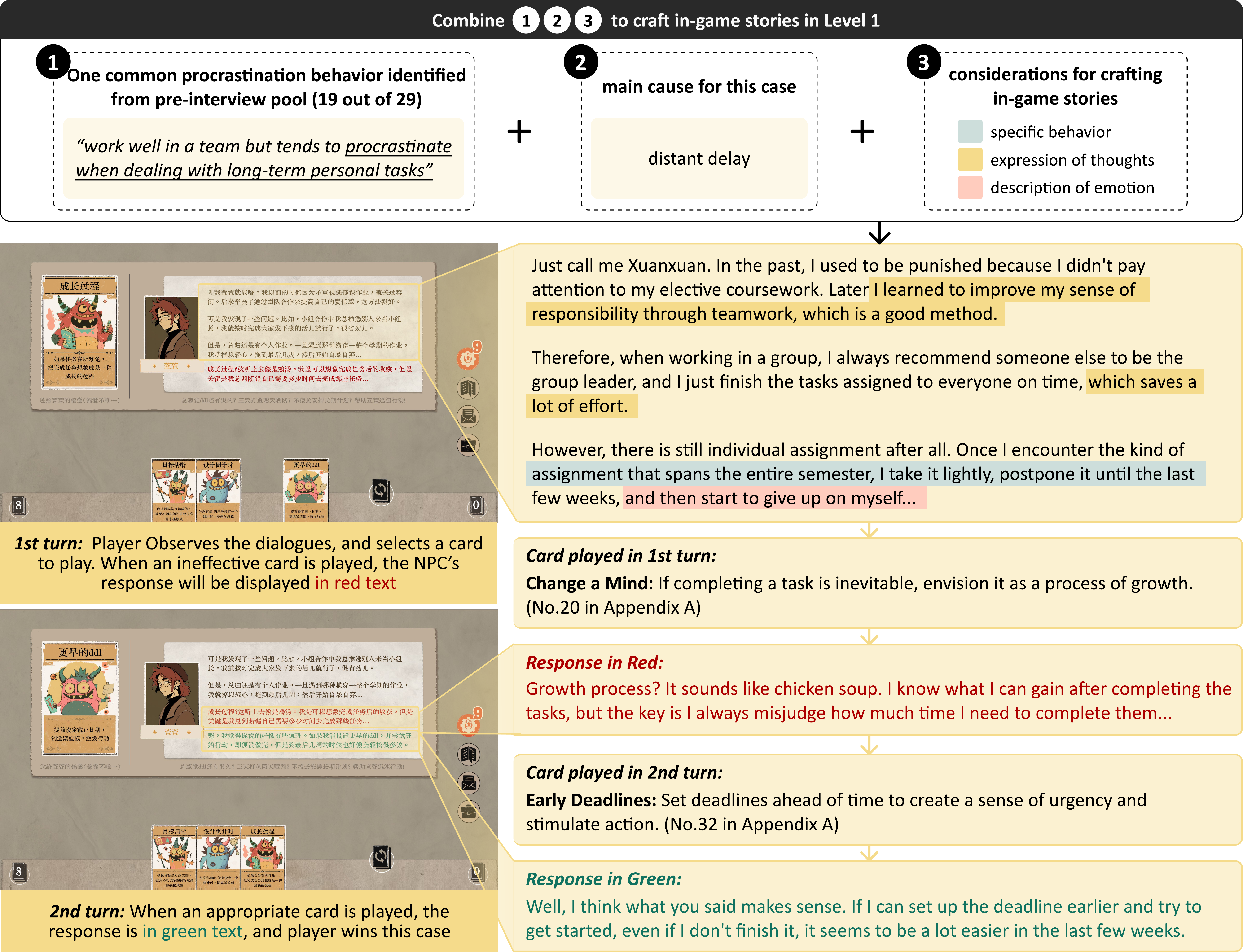}
  \caption{Upper: Three design considerations for crafting procrastination stories in Level 1. Lower: A translated, detailed example of core gameplay in Level 1. }
  \label{level1}
\end{figure*}

\section{Design and Implementation of \textit{ProcrastiMate}}
\label{procrastimate section}
\hl{Following the formative study, this section aims to address our second research objective. Specifically, we developed a text-based adventure game called \textit{ProcrastiMate}, incorporating three design considerations aligned with design goals: 1) customizing in-game stories, 2) casting players as a counselor, and 3) providing adaptive dialogues.}

\subsection{Game Setting}
\subsubsection{Story Line}
\textit{ProcrastiMate} is set in a \hl{fictional} university. Players' primary objective is to transform the institution's traditional punitive approaches to managing procrastination into a more supportive environment. Specifically, players work to replace the \textsc{Punishment Handbook} and establish a \textsc{Management Handbook}. The \textsc{Punishment Handbook} lists four common misconceptions about procrastination: \textsc{Incompetence}, \textsc{Irresponsibility}, \textsc{Weak Willpower}, and \textsc{Laziness}. In contrast, the \textsc{Management Handbook} contains four chapters offering strategies to address the causes: \textsc{Improve Self-efficacy}, \textsc{Embrace Task Value}, \textsc{Control Impulsiveness}, and \textsc{Adjust Distant Delay}. \hl{To complete the \textsc{Management Handbook}, players must populate each chapter with six coping strategy cards, requiring players to solve 24 (4 * 6) in-game cases. As shown in \autoref{fig:gameplay}-a, this structure is grounded in Steel's theoretical framework \cite{steel2007nature} as well as the 40 coping strategies (see \autoref{appendix_card}) we identified from literature.}

\subsubsection{Players Take on the Role of A Counselor}
In \textit{ProcrastiMate}, players assume the role of a counselor, using coping strategy cards to help in-game characters address their procrastination issues. By framing player role in this way, we aimed to minimize the negative emotional responses reported by participants in the formative study, while creating the necessary psychological distance for effective learning.

\subsubsection{Customised Procrastination Stories}
\label{craft_stories}
\hl{To ensure the game content resonated with players' real-life experiences, we customized the procrastination stories in Levels 1 and 2 of \textit{ProcrastiMate}. For Level 1, we created a shared pool of 24 stories based on an analysis of scenarios described by participants (N = 29) during pre-interviews (\autoref{evaluation section}). These stories reflected four key clusters of procrastination behavior among college students: daily routines (e.g., student services), study-related tasks (e.g., assignments, exams, research projects), health and fitness, and self-improvement activities (e.g., learning new skills). Designed with the considerations outlined in \autoref{level1}, 24 cases in Level 1 intended to align with players' college-related contexts.}

\hl{For Level 2, we adopted an individualized customization approach with considerations outlined in \autoref{level2}, tailoring the stories to each player's unique procrastination experiences shared during pre-interviews. As a result, each participant encountered a different set of Level 2 cases\footnote{\hl{The customization for Level 2 was implemented by assigning each player a unique \texttt{.json} file containing their personalized stories within the game. The game, tailored with these files, was then provided to each participant. For participants who shared fewer than eight procrastination behaviors (Level 2's total cases), the missing cases were supplemented randomly from other participants' stories.}}. This customization aimed to provide a personalized narrative in Level 2, reflecting each player’s procrastination patterns as the game progressed.}

\subsection{Level 0: Establishing the Foundation of Procrastination Understanding}
Before progressing to Levels 1 and 2, Level 0 serves as an introductory level where players get familiar with \textit{ProcrastiMate} concept. In this level, players engage with eight historical procrastination cases that were previously misunderstood and harshly penalized based on an outdated handbook. The primary objective for players is to identify the specific support the NPCs actually need, rather the harsh criticisms. 

For each case, players review: 1) NPC's basic information, 2) case description, 3) the label of misunderstanding, and 4) the assigned punishment. Players must then choose one of four types of suggestions: \textsc{Improve Self-efficacy}, \textsc{Embrace Task Value}, \textsc{Control Impulsiveness}, or \textsc{Adjust Distant Delay}. These options directly correspond to the four chapters of \textsc{Management Handbook} that players will develop in Level 1.

\subsection{Level 1: Building Connections Between Coping Strategies and Procrastination Causes}
After the introductory warm-up in Level 0, players start to develop the \textsc{Management Handbook} in Level 1. As introduced in \autoref{fig:gameplay}-a, the handbook is organized into four chapters, each comprising six cases. As players navigate these cases, they \textbf{\textit{experiment with which strategy cards effectively counter specific causes of procrastination}}, ultimately populating the handbook with 24 (4 * 6) validated strategies. This design grants players a certain degree of freedom to choose their approach, selecting 24 out of 40 available strategy cards.

\textbf{Gameplay in Each Case}: At the start of Level 1, players are assigned 16 strategy cards. As illustrated in \autoref{fig:gameplay}-b, each NPC case is labeled with a major cause, aligning with the chapter's theme. Players engage with the NPCs through dialogues to understand each case and can experiment with different strategy cards. \textbf{The win condition} is determined by matching appropriate coping strategies to their corresponding causes. For instance, cards No.1 to No.10 were designed for \textsc{Improve Self-efficacy}. If a player encounters a procrastination scenario caused by low self-efficacy, using any card from this range can secure a win. This gameplay design encourages exploration, allowing multiple correct solutions within each cause category and motivating players to try various strategies.

During the game, players will receive immediate \textbf{adaptive dialogue feedback} on their choices through NPC responses. A translated, detailed example in one case was presented in \autoref{level1}. Specifically, ineffective cards prompt critical feedback from the NPC displayed in red text, explaining why the strategy does not work for that case. While successful strategies elicit positive responses displayed in green text. This feedback system helps players understand why an NPC accepts or rejects a particular coping strategy (card played). 

Successfully resolving a case automatically adds the employed strategy card to its corresponding chapter in the \textsc{Management Handbook} and rewards the player with a \textbf{Privilege Point}, which can be used to acquire new strategy cards. The card acquisition system is carefully balanced. As players complete the four chapters, they can gain an additional 24 cards through dialogue interactions with NPCs (using Privilege Points) or from thank-you letters sent by NPCs. By the end of Level 1, players will have a total of 40 cards.
A chapter is completed when all six cases are successfully addressed, meaning six effective cards are placed within it. This process of solving problems and building the handbook forms the core gameplay loop. Players \textbf{win Level 1} when all four chapters are completed.

\begin{figure*}[t]
  \centering
  \includegraphics[width=\linewidth]{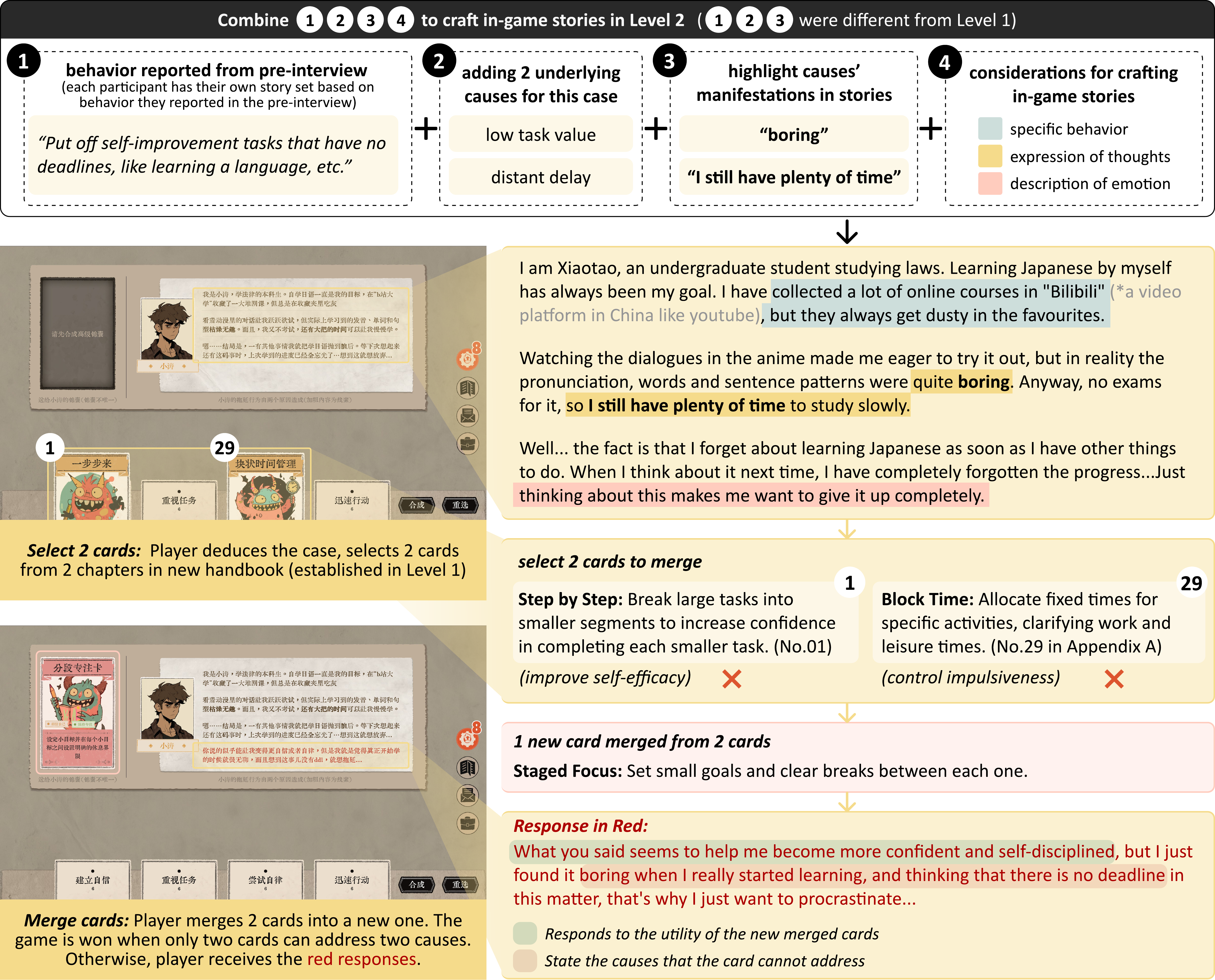}
  \caption{Upper: Four design considerations for crafting procrastination stories in Level 2. Lower: A translated, detailed example of core gameplay in Level 2.}
  \label{level2}
\end{figure*}

\subsection{Level 2: Create Personalized Coping Strategies for Complex Cases}
After completing Level 1, players are expected to have developed a solid understanding of the causes of procrastination and the corresponding coping strategies. Building on this foundation, players now use the 24 strategy cards they have collected in the new handbook to tackle eight more complex cases.

The gameplay in this final level of \textit{ProcrastiMate} is similar to that of Level 1 (see \autoref{fig:gameplay}-c); however, the challenge is heightened by requiring players to \textit{\textbf{diagnose two underlying causes for each procrastination case}}. Players must select two cards from their existing handbook and combine them to create a new, merged coping strategy card. Success is determined by whether the player uses the correct strategies to address both causes. For example, if a case is caused by both low self-efficacy and low task value, the player must combine one card from \textsc{Improve Self-efficacy} and another from \textsc{Embrace Task Value} for winning. Players have the freedom to merge and update the coping strategies they choose for each case, and the final merged cards reflect the players' personalized choices. An example of gameplay in this level is shown in \autoref{level2}.

\begin{figure*}[t]
  \centering
  \includegraphics[width=\linewidth]{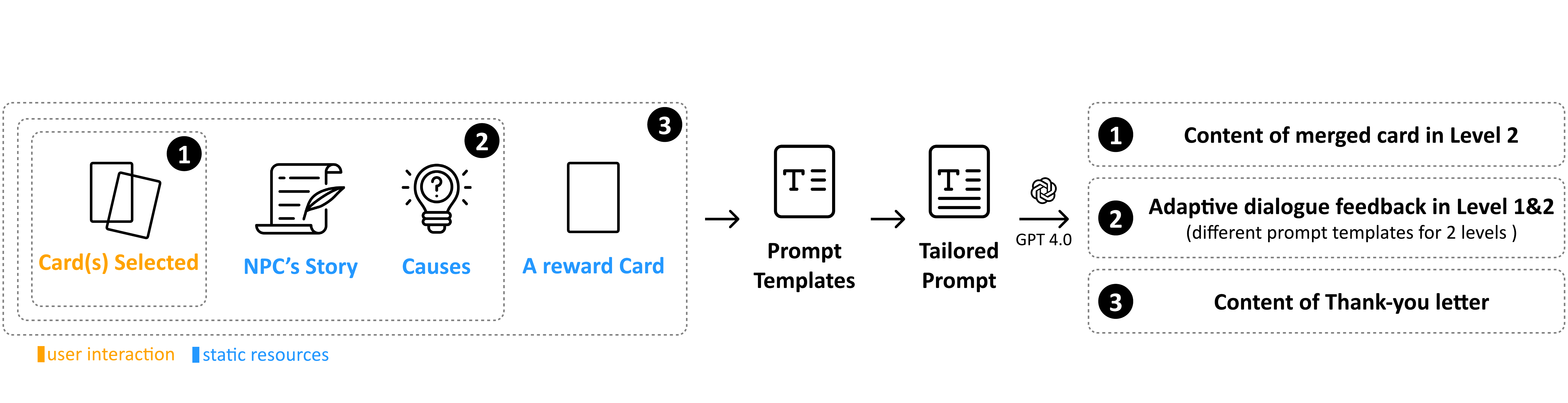}
  \caption{Overview of OpenAI API calls in \textit{ProcrastiMate} (see supplementary for predefined prompt templates).}
  \label{prompt}
\end{figure*}

\subsection{Implementation}
\textit{ProcrastiMate} was developed using the Unity engine, with versions available for both MacOS and Windows platforms. To facilitate the field study evaluation, we implemented an auto-save feature that captures and stores the player's progress locally in JSON format upon exit, ensuring seamless continuity across gaming sessions. Furthermore, we integrated OpenAI's GPT-4.0 API into \textit{ProcrastiMate} to generate adaptive dialogue feedback, thank-you letters, and the content of merged cards in Level 2. \autoref{prompt} depicts the game produces tailored prompts by combining various static resources and real-time user actions with predefined prompt templates.

\section{Evaluation Methods of \textit{ProcrastiMate}}
\label{evaluation section}
\hl{To address our third research objective, we evaluated \textit{ProcrastiMate} by conducting a field study within 2 weeks with 27 participants. In this section, we detailed our evaluation methods, which was guided by two research questions: }
\begin{itemize}
    \item \textbf{RQ1}: How does \textit{ProcrastiMate} support players in learning about procrastination's causes and coping strategies?
    \item \textbf{RQ2}: How do the specific design considerations (based on our design goals) of \textit{ProcrastiMate} shape the user experience and facilitate the learning processes?
\end{itemize}

\hl{Specifically, we first introduced the participants recruitment. Second, we presented the procedures of our field study as well as the quantitative and qualitative data collection. Finally, we detailed how we analyze the data collected. The evaluation study received the ethical approval from the first author's institution. }

\subsection{Participants}
Recruitment was conducted through an open call advertisement posted on public social media platforms at a university in China. Participants were required to self-identify as experiencing life challenges related to procrastination. Individuals who had received psychological treatment or had been diagnosed with a mental disorder in the past three months were excluded from the study (N = 21). A total of 29 participants were recruited and completed the pre-interview. During the study, two participants did not respond to follow-up emails for the post-interview, leaving 27 participants (12 male, 15 female; other gender options were also provided) who completed the entire study. Before the pre-interview, we collected participants' demographic information (see \autoref{Demographics}) and assessed their procrastination tendencies using the General Procrastination Scale \cite{lodha2019general}. Each participant received a bonus of 80 CNY after completing the study. 

\begin{table}
  \caption{Demographic information of 27 participants}
  \label{Demographics}
  \resizebox{\linewidth}{!}{
  \begin{tabular}{ll}
    \toprule
    ~ & Total Participants (N = 27) \\
    \midrule
    \textbf{Gender} & Male(44\%), Female(56\%)\\
    \textbf{Age} & 19-29 (M = 23.26, SD = 2.35)\\
    \textbf{Education} & \makecell[l]{Undergraduate(56\%), Postgraduate(26\%),\\ PhD(15\%), Prefer not to say(3\%)}\\
    \textbf{Background} & \makecell[l]{Design/Art(30\%), Engineering(18\%),\\
    Natural Sciences(7\%), Social Sciences(7\%),\\
    Humanities(7\%), Marine Science(4\%),\\
    Law(4\%), Education(4\%), \\
    Computer Science/Information Technology(4\%), \\
    Business/Management(4\%), Prefer not to say(11\%)}\\
    \textbf{\makecell[l]{Trait Procrastination\\ Tendencies}} & \makecell[l]{High(63\%), Above Average(19\%), \\Average(7\%), Low(11\%)}\\ 
  \bottomrule
\end{tabular}}
\end{table}

\subsection{\hl{Procedures and Data Collection}}
\hl{Our evaluation study involved three main stages, incorporating both qualitative and quantitative data collected. Specifically, we conducted semi-structured interviews to investigate RQs, assessed participants' emotional attitudes toward procrastination in both pre- and post-study, and measured gaming and learning experience in post-study to provide complementary indicators of user experiences.} 

\hl{\textbf{Stage 1 - Pre-Interview}: Interviews were conducted either online\footnote{\hl{For both pre- and post-interviews conducted online, the Tencent Meeting platform, one of the most popular videotelephony platforms in the mainland of China, was used.}} or offline, based on participants' availability. Participants were introduced to the study and consent was obtained, followed by the collection of data:}
\begin{itemize}
    \item \hl{\textit{Quantitative}: Participants completed the Negative and Positive Emotions Attributed to Procrastination Scale (NPEAPS) \cite{lee2020understanding} on a 5-point Likert scale to assess procrastination-related emotions.}
    \item \hl{\textit{Qualitative}: A semi-structured interview was then conducted, focusing on participants' experiences with procrastination behavior, their coping strategies and attitudes toward procrastination (see Supplementary materials for full interview outline). The interviews were audio-recorded.}
\end{itemize}

\hl{\textbf{Stage 2 - Field Study}: Participants engaged with the \textit{ProcrastiMate} game during this stage. Insights from the pre-interviews were utilized to tailor the in-game procrastination stories (as described in \autoref{craft_stories}), which were pilot tested with 10 external participants to ensure high relatability. Subsequently, \textit{ProcrastiMate}, available for both MacOS and Windows, was distributed along with installation instructions. Participants played the game at their own pace over a two-week period. An in-game notification prompted them to contact the research team upon reaching the midpoint of Level 2, encouraging them to schedule a post-interview session. Those who had not scheduled an interview by the tenth day received a reminder.}

\hl{\textbf{Stage 3 - Post-Interview}: This stage was designed to gather in-depth insights into the participants' experiences with the game and their learning outcomes. Conducted similarly to Stage 1, the post-interviews were either online or offline, depending on participant availability. The data collected at this stage included:} 
\begin{itemize}
    \item \hl{\textit{Quantitative}:}
    \begin{itemize}
        \item \hl{\textit{Procrastination-related emotions}: 
        Measured by completing the NPEAPS again \cite{lee2020understanding}.}
        \item \hl{\textit{Engagement in Learning}: Measured using the Experiencing Scale (Long Form) on a 7-point Likert scale \cite{stock2021experiencing}.}
        \item \hl{\textit{Enjoyment of game}: Measured using the EGameFlow Questionnaire on a 7-point Likert scale \cite{shu2018validation}.}
        \item \hl{\textit{Empathy for in-game characters}: Measured using the Scale of State Empathy on a 5-point Likert scale \cite{shen2010scale}.}
    \end{itemize}
    \item \hl{\textit{Qualitative}: The post-interview concluded with a semi-structured interview about participants' understanding of procrastination based on the game content and their feedback on the gaming experience (see Supplementary materials for full interview outline). The interviews were audio-recorded.}
\end{itemize}

\subsection{\hl{Data Analysis}}
\subsubsection{\hl{Quantitative data analysis.}} \hl{First, we checked the normality of the data collected by NPEASP using the Shapiro-Wilk test and calculated Cronbach's alpha to assess the reliability of the scales. We then conducted paired-sample t-tests to compare pre- and post-test scores on the NPEAPS to investigate whether participants showed changes in emotions related to procrastination. For the other scales, we calculated descriptive statistics (mean and standard deviation) to assess participants' engagement in the learning process, enjoyment, and empathy, with the median score serving as a benchmark for evaluation. The results were presented in \autoref{quant}}

\subsubsection{\hl{Qualitative data analysis.}}
\hl{The pre-interviews, as mentioned earlier, were used to identify common procrastination scenarios among college students. The post-interviews generated approximately 16.5 hours of audio recordings, which were transcribed verbatim to create the dataset for investigating our RQs.  Following Braun and Clarke's thematic analysis procedures ~\cite{braun2006using}, we adopted an integrated deductive and inductive approach. Three authors served as coders and collaboratively conducted the analysis pipelines for the RQs.}

\hl{\textbf{For RQ1—investigating the learning processes and outcomes}, our deductive analysis was guided by the Experiential Learning Cycle (ELC) model proposed by David Kolb~\cite{kolb2014experiential}. The ELC is a well-established framework for understanding how experience, reflection, and action foster learning and has been previously utilized in HCI research~\cite{ibrahim2024supporting, saksono2019social}. The model outlines four key stages: 1) \textbf{Concrete Experience (CE)}—actively engaging in an activity; 2) \textbf{Reflective Observation (RO)}—reviewing and reflecting on the experience; 3) \textbf{Abstract Conceptualization (AC)}—drawing conclusions and developing new insights; and 4) \textbf{Active Experimentation (AE)}—applying new understanding to future situations. Using the ELC framework, we aimed to explore users’ learning processes during interaction with \textit{ProcrastiMate}. The inductive portion of the analysis allowed new themes or sub-themes to emerge directly from the data.}

\hl{Specifically, three coders independently familiarized themselves with the dataset and generated initial codes under the ELC framework. Through iterative discussions, the coders refined the codes and reached a consensus, particularly clarifying distinctions between RO and AC stages. We agreed that, in our context, RO should include codes and quotes about reflection on personal experiences, such as self-identified procrastination triggers, while AC should reflect broader insights into procrastination. The final shared codebook comprised 12 sub-themes from 40 codes, such as “\textit{personal procrastination triggers}” and “\textit{emotions regarding procrastination}”. Example codes included “\textit{new coping strategy learned}”, and “\textit{link to personal life}”. The findings for RQ1 were reported in \autoref{finding1}. }

\hl{\textbf{For RQ2—investigating learning experiences}, the deductive analysis examined three design considerations: 1) customizing in-game stories, 2) casting players as a counselor, and 3) providing adaptive dialogues, to understand their impact on learning experiences. The inductive approach allowed us to identify additional themes or sub-themes that emerged from the data. Following a similar pipeline as RQ1, the analysis identified six sub-themes from 28 codes, such as “\textit{eliciting empathy}” and “\textit{more objective judgment}”. Example codes included “\textit{personal pronouns}”, “\textit{red dialogue feedback}”, and “\textit{sense of immersion}”. The findings for RQ2 were reported in \autoref{finding2}. Additionally, a new theme, titled “\textit{tensions and challenges}”, emerged from the inductive analysis, highlighting the challenges in the educational approach. These findings were detailed in \autoref{finding3}.}

\section{Findings}
\label{findings section}
\hl{In this section, we present findings from our evaluation, addressing our RQs and fulfilling our third research objective. We begin with results from the quantitative analysis (\autoref{quant}) that offered supplementary insights into participants' learning experiences, then transition to qualitative findings addressing RQ1 (\autoref{finding1}), RQ2 (\autoref{finding2}), and identified tensions and challenges (\autoref{finding3}).}

\subsection{Quantitative Findings}
\label{quant}
Our quantitative results suggested the gaming experience led to \textit{a decrease in negative emotions} associated with procrastination. Specifically, the paired-sample t-test showed a significant reduction in negative emotions post-test (M = 2.79, SD = 0.69) compared to pre-test (M = 3.19, SD = 0.82), t (26) = 2.869, p < 0.01. However, no significant change was observed in positive emotions, with post-test scores (M = 2.09, SD = 0.71) remaining statistically similar to pre-test scores (M = 1.93, SD = 0.89), t (26) = 1.131, p = 0.269. In addition, descriptive statistics for the other three scales revealed \textit{generally positive experiences} across all dimensions. The mean score on the Experiencing Scale was 5.80 (SD = 0.84), with most participants scoring above the median, suggesting that the game fostered meaningful engagement and learning. The EGameFlow Questionnaire yielded a mean score of 5.57 (SD = 0.85), indicating high levels of immersion and satisfaction with the game. On the Scale of State Empathy, measured on a five-point scale, the mean score was 4.30 (SD = 0.67), reflecting a positive trend. The results are detailed in Table \ref{Descriptive Statistics}.

\begin{table}[t]
    \caption{The results of the Experiencing Scale, EGameFlow and Scale of State Empathy}
    \resizebox{\linewidth}{!}{
    \label{Descriptive Statistics}
    \renewcommand\arraystretch{1}
    \begin{tabular}{cccc}
    \toprule
        \textbf{Scale} & \textbf{Dimensions} & \textbf{M} & \textbf{SD} \\ 
        \midrule
        \textbf{\multirow {4}{*}{\makecell[c]{The Experiencing Learning \\Scale (Long Form)}} }
        & Novelty & 5.96 & .84 \\ 
        \textbf{} & Presence & 5.75 & .87 \\ 
        \textbf{} & Embodiment & 5.85 & 1.11 \\ 
        \cmidrule(r){3-4}
        \textbf{} & ~ & 5.80 & .84 \\ 
        \midrule
        \textbf{\multirow {7}{*}{EGameFlow} }
        & Concentration & 5.65 & 1.01 \\ 
        \textbf{} & Goal Clarity & 5.73 & .71 \\ 
        \textbf{} & Feedback & 5.56 & 1.39 \\ 
        \textbf{} & Challenge & 5.59 & 1.06 \\ 
        \textbf{} & Autonomy & 5.54 & 1.13 \\ 
        \textbf{} & Immersion & 5.34 & .98 \\ 
        \cmidrule(r){3-4}
        \textbf{} & ~ & 5.57 & .85 \\ 
        \midrule
        \textbf{\multirow{4}{*}{Scale of State Empathy}} 
        & Affective Empathy & 4.22 & .76 \\ 
        \textbf{} & Cognitive Empathy & 4.43 & .70 \\ 
        \textbf{} & Associative Empathy & 4.26 & .71 \\ 
        \cmidrule(r){3-4}
        \textbf{} & ~ & 4.30 & .67 \\
        \bottomrule
    \end{tabular}}
\end{table}

\subsection{Qualitative Finding 1 (RQ1): Understanding How \textit{ProcrastiMate} Supported Learning About Procrastination}
\label{finding1}

\hl{To address RQ1, in this section, we present how \textit{ProcrastiMate} facilitated players' learning about procrastination by organizing the findings into four themes corresponding to the four stages of the ELC. }

\subsubsection{Concrete, Emotionally Engaging Experiences for Learning About Procrastination}
According to Kolb, concrete experience is the foundation of the learning process. In \textit{ProcrastiMate}, players were provided with concrete, emotionally engaging experiences to learn about procrastination through empathetic storytelling, a low-risk environment, and the impactful decision-making process. 

\hl{First, we found that \textbf{emotionally detailed stories enhanced players' engagement with the game}. In \textit{ProcrastiMate}, the diverse procrastination stories went beyond merely describing behaviors. They also incorporated the emotions, thoughts, and rationalizations of NPCs.} Participants found that \textit{“these details make the NPCs' story more realistic”} (P01), which encouraged empathy toward the NPCs and prompted them to approach the stories with greater considerations. For example, P06 mentioned: \textit{“An NPC felt his paper was not thorough enough and still needed more materials or experiments, which was why his progress was delayed. So, he did not think he was procrastinating. I somehow agreed with his thoughts and feelings, so I was more careful on that case.”}

\hl{Second, the absence of harsh punishments in \textit{ProcrastiMate} created \textbf{a low-risk environment that encouraged players to explore and engage with more game content}.} For example, P25 shared: \textit{“There were no harsh punishments in the game, so I felt comfortable trying different coping strategies for each case, just to see what would happen.”}

\hl{Third, by tasking players with diagnosing the causes of procrastination and selecting appropriate coping strategies, the gaming process involved an interactive simulation of procrastination management that\textbf{ encouraged active thinking, which participants found enjoyable}.} As P13 shared, \textit{“the time and effort spent on the game was meaningful because it really made me think”}. Players like P10 expressed similar appreciation, noting \textit{“I never expected a game to help me understand procrastination. I liked the dynamic and interactive gaming process. Before, I used to watch videos on overcoming procrastination online, but they were not very fun.”}

\subsubsection{Self-Reflection on Procrastination Causes, Strategies, and Attitudes}
Based on the concrete, enjoyable experiences, we found \textit{ProcrastiMate} fostered players' reflective observation by encouraging them to think critically about their own procrastination behaviors, coping strategies, and preconceived notions about procrastination. 

\hl{First, as \textit{ProcrastiMate} categorized all NPCs' stories according to Piers Steel's four causes of procrastination, it provided players with\textbf{ new perspectives to reflect on their own procrastination triggers}.} As P19 noted, they \textit{“gained a structured approach to ask the ‘why’ behind procrastination behavior”}. Several participants reported that they began analyzing their own procrastination triggers using these four causes. For example, P17 reflected: \textit{“I started using the four causes to think about why I procrastinate on my master’s thesis proposal. First, I believe I can handle it. Second, I definitely value it highly—otherwise, I would not be able to graduate... I think the lack of urgency is probably the most important reason. Since there are still two months left, I feel like this task does not require all that time to complete.”}

\hl{Besides, as the game required players to continuously experiment with various coping strategies, it\textbf{ encouraged players to reflect on the effectiveness of their own strategies}}. For example, P23 shared she used to reward herself with meals at fancy restaurants for completing tasks early, but after playing the game, she reconsidered this approach: \textit{“Now when I think about that strategy again, I realize it's a way to increase task enjoyment and boost motivation.”} Similarly, the game encouraged players to rethink strategies that may have been less effective. P19, for instance, often made detailed plans, but when she applied similar strategies for the NPCs and received negative feedback, she reflected on how this approach might actually be a way to avoid the real work: \textit{“After making a plan, I always felt like I had already accomplished a lot and could rest. But in reality, it did not boost my confidence or change my attitude toward the task.”}

\hl{Third, we found that the narrative in \textit{ProcrastiMate}, shifting from punitive styles to more supportive approaches,\textbf{ encouraged players to question their stereotypes about procrastination}}. For instance, some participants had previously viewed procrastination as purely a result of laziness or a lack of discipline. P17 reflected on this shift in perspective: \textit{“I used to accuse people who procrastinated more in group projects of being irresponsible. But now I think it is possible that they just do not like the task they were assigned, or maybe their task is too difficult.”} (P17)

\subsubsection{New Insights and Reduced Negative Emotions About Procrastination Through Structured Learning}
Building on the concrete experiences and reflective observation stages, we found that \textit{ProcrastiMate} also guided players toward abstract conceptualization. For example, players began to see procrastination as a structured problem, a process-driven challenge, and an emotionally nuanced experience, rather than just isolated behaviors.

\hl{First, the game's structured approach to procrastination management became evident through its organization into four chapters, each containing six strategy cards. This layout helped players organize the scattered coping strategies, \textbf{fostering a more structured framework of procrastination management}.}. As P16 summarized: \textit{“First, identifying the cause, and then selecting the appropriate coping strategies.”} This systematic and actionable approach was further emphasized by P22, who noted: \textit{“This game made me realize that procrastination can be understood in terms of these few causes. Once I identify a cause, there are corresponding strategies to use. This structured approach gives me a clearer way of thinking about procrastination.”} In addition, this organized layout was seen as \textit{“quite conducive to memory”} (P27), as it helped players retain key strategies more effectively.

\hl{Besides, in \textit{ProcrastiMate}, several NPCs were designed to seek help from players repeatedly for different reasons, which \textbf{encouraged players to develop a process-oriented view of overcoming procrastination}}. As P21 noted, \textit{“Some NPCs would return, which made me realize that treating procrastination is not necessarily a root-and-branch solution, but a process of constantly practicing and optimizing methods.”} Additionally, Level 2 introduced more complex procrastination cases, helping players understand that procrastination can be driven by a combination of causes that vary depending on the \textit{“situation and individual”} (P12). 

\hl{Lastly, our analysis found that learning about the causes and coping strategies helped players \textbf{foster a more compassionate and objective view of procrastination}}. P03 expressed this sentiment: \textit{“With these causes as powerful tools, I think I will be able to view my procrastination more objectively. This will make me feel less guilty, as I can now explain why I do it.”} Additionally, by conveying that procrastination has psychological underpinnings and is not inherently negative, the game helped alleviate the guilt and self-blame often associated with it. As P15 shared: \textit{“I used to feel bad when I procrastinated, which would actually lead to more anxiety and procrastination—a vicious cycle. But the game reminds me that procrastination is a normal phenomenon in humans, with common causes shared by many.”}

\subsubsection{Real-World Experimentation Beyond Game Environment}
\hl{Through engaging with \textit{ProcrastiMate}, players also reported that they applied the knowledge and coping strategies learned in the game to their real-life situations.}

\hl{First, the abundance of coping strategies introduced in the game made players feel as if they were \textit{“opening a treasure chest, realizing there were so many strategies to choose from”} (P17). Several participants described how they \textbf{explored these newly learned coping strategies in their daily experiences}}. For instance, P17 shared how she experimented with strategy No. 28, \textit{Record and reflect on your work habits and factors that distract your attention}: \textit{“I did not know about this method for analyzing procrastination before. But when I received psychological counseling in the past, the counselor suggested a similar approach to analyze emotional causes and find solutions. I think it is very similar to the process described in this strategy. Recently, I have been applying it to my own procrastination behavior.”}

\hl{Moreover, players also reported that they consciously or unconsciously \textbf{combined different strategies from the game to create customized approaches tailored to their specific circumstances}}. For example, P25 described how various strategies influenced his work habits: \textit{“Just yesterday, the project leader asked me to draw some sketches. Normally, I would have procrastinated until the evening because I felt I could pull an all-nighter. However, the strategy ‘earlier deadlines’ (card No.32) struck me, and with that awareness, I completed the task in the afternoon. I also thought about how much more relaxed I would feel after finishing it (card No.17)”} .

\hl{Lastly, an interesting outcome of our analysis was players' \textbf{desire to incorporate game elements into their daily routines in tangible formats.}} For instance, P10 shared: \textit{“I really like creating handbooks, and sometimes I add my own plans to them. I want to turn these cute cards into physical ones and maybe even draw one every day, like tarot cards, to motivate myself to complete daily tasks.”} Similarly, P23 mentioned saving screenshots of the game's strategies for easy reference: \textit{“I took screenshots of the four chapters in Level 1 and the corresponding coping strategies I built, and put them on my desktop. They are ready to use whenever I realize I am procrastinating.”} These examples suggest that the effects of the game may extend beyond the virtual environment, potentially influencing players' real-life behavior on a daily basis.

\subsection{Qualitative Finding 2 (RQ2): Contextual Examination of Our Design Considerations}
\label{finding2}
\hl{To address RQ2, in this section, we present the  findings on how our three design considerations facilitated the learning processes and shaped the gaming experience. }

\subsubsection{‘That's Exactly Me!’: Relatability and Reflection Facilitated by Customized Narratives}

\hl{Firstly, the in-game customized procrastination stories struck a chord with players, \textbf{creating a strong sense of relatability and eliciting empathy}}. For example, P12 noted: \textit{“The aircraft cabin design project (in Level 2), where the teacher’s requirements seemed too detailed and impossible to handle, mirrored my own experiences exactly.”} This relatability elicited emotional responses, with another player stating, \textit{“Many stories in the game feel like they are about myself”} (P06). The familiar and relatable narratives also led to more thoughtful selection of strategy cards. P21 noted, \textit{“It felt like my friends were seeking help from me, so I took it quite seriously”}. 

\hl{Moreover, these relatable cases \textbf{bridged players' in-game experiences with their real-life experiences, prompting actual reflections and actions}}. For example, P15 reflected on his procrastination with language learning: \textit{“The German learning scenario made me think about my Japanese studies. Initially, the syllabaries were easy, but I have been avoiding grammar. The game suggested increasing interest and removing distractions, which made me realize, ‘aha, I do find grammar boring.’ This relevance helped me understand why I delay learning languages.”} Similarly, P30 rethought his approach to staying motivated in sports, sharing: \textit{“I am a professional student-athlete, but like the example in the game says, sports can be boring sometimes. Helping the NPCs made me realize how these strategies can make exercise more enjoyable.”} These reflections demonstrate how the relatable scenarios encouraged players to apply the knowledge directly to their own lives.

\subsubsection{‘The Bystander Sees More Clearly’: Enhancing Objective Learning and Psychological Safety Through the Advisor's Role}

\hl{Our evaluation found that positioning players as advisors \textbf{allowed them to analyze procrastination scenarios more objectively}. A majority of participants (22 out of 27) reported that assisting in-game characters helped them adopt a more open-minded approach, considering a broader range of strategies beyond their personal biases}. P20 remarked, \textit{“Helping others gives me a god-like perspective. I consider what choices would be beneficial for them and make decisions based on their goals.”} This method also helped players avoid self-bias when addressing procrastination issues. Another participant noted, \textit{“It is easier to see clearly why others procrastinate. When it comes to myself, I tend to make excuses and emotional decisions, just like the NPCs.”}

\hl{On the other hand, we found that interacting with NPCs to discuss NPCs' procrastination issues \textbf{effectively reduced feelings of guilt or defensiveness} among players}. P27 mentioned, \textit{“I am less resistant to engaging with and understanding the game due to the indirect approach.”} This design strategy created a safe psychological distance, minimizing the defensive reactions often triggered by self-reflection on sensitive topics. P16 shared, \textit{“Directly discussing my own procrastination usually feels like being judged, which will be quite unpleasant.”} Similarly, P28 observed, \textit{“This approach makes me less tense and more receptive to new knowledge.”}

\subsubsection{‘It Seems like the NPC is having a discussion with me’: Smoothing Knowledge Understanding Through Adaptive Dialogues}

\hl{The integration of a responsive dialogue system was crucial for \textbf{aiding players in grasping in-game concepts and strategies}}. When a player selected an ineffective strategy card, \textit{“... the NPC’s feedback clarified why the choice did not work and prompted them to reconsider my strategies”}(P19). Furthermore, NPCs' feedback was crucial in helping players embrace the game’s underlying logic. P11 described their experience: \textit{“Initially, I tried to guess the correct answers based on my personal experience, but the dialogue guided me to understand the game’s intended approach.”}

\hl{Besides, the interactive dialogue system also \textbf{boosted player immersion} by simulating realistic conversations with NPCs}. It evoked emotional investment, making players feel \textit{“genuinely involved in assisting the NPCs”} (P04). P21 described the impact of this feature on their engagement: \textit{“The guidance in red text and the NPC’s responses make it feel like I am having a real discussion with them. Sometimes the NPC disagrees with my advice, which adds to the realism and depth of our interaction.”}

\subsection{\hl{Qualitative Finding 3: Challenges and Tensions Observed}}
\label{finding3}
\hl{The above mentioned findings demonstrated how our approaches supported participants in learning about procrastination with an engaging and safe gaming experience. However, during our analysis, several tensions and challenges also emerged, regarding how participants engaged with and interpreted the game's educational content.}

\hl{Firstly, an unexpected challenge emerged when two participants expressed \textbf{concerns that increased understanding of procrastination mechanisms could potentially enable the justification of the behavior}. For example, P7 noted, \textit{“Understanding these causes made me feel relieved, but I am a little worried that they might become an excuse for me to rationalize my procrastination.”}}

\hl{Besides, our analysis revealed \textbf{tensions between personal definitions of procrastination and the game's theoretical framework}. Based on Steel's work, the game presented a few procrastination stories stemming from perceived low task value. However, some participants did not consider avoiding unimportant tasks as procrastination. P22 shared, \textit{“I think procrastination is when there is something beneficial to you, but you keep putting it off... If it is something meaningless that you do not want to do, you might as well not do it - I do not think that is procrastination.”}. This tension would further affected their engagement with the game, as P22 explained: \textit{“When dealing with the case where NPC was putting off the assignments from mandatory but irrelevant courses, I truly did not find a suitable coping strategy as I did not think it was procrastination.”}}

\hl{Another \textbf{tension emerged between knowledge acquisition and expectations for behavioral change}. While our game facilitated players' understanding, reflection, and experimentation with newly gained knowledge, two participants raised concerns about the potential for sustained behavioral change. Although behavior change was not the direct aim of our current study, it is essential to recognize the participants' desire for an integrated approach that combines understanding with practical tools for behavioral change. P10 expressed a need for more actionable guidance: \textit{“I wish the game provided more guidance on how to apply this knowledge in daily life and encouraged me to keep doing it consistently.”} Similarly, P2 noted, \textit{“I understand this is to help me learn more about procrastination-related knowledge, but I would prefer to have it combined with direct intervention in my behavior.”}}

\section{Discussion}
\hl{The above-mentioned findings presented three key aspects of \textit{ProcrastiMate}'s educational approach: its effectiveness in supporting procrastination learning, the role of specific design elements in shaping user experiences to facilitate effective learning, and critical insights from users about this educational approach.} 

\hl{Correspondingly, we examined the broader implications of these findings in this section. First, we discuss how our educational perspective, using serious games as a medium, contributes to the ongoing discourse in HCI, both within and beyond the domain of procrastination interventions. Next, we reflect how \textit{ProcrastiMate} fosters effective learning by maintaining psychological distance from players, offering an alternative perspective for future serious game design. Third, we reflect on the challenges observed in our study and the implications for future work. Finally, we acknowledge the limitations of our current work.}

\subsection{Breaking the Procrastination Cycle with Game-based Educational Approaches}

\hl{Previous HCI research on procrastination has predominantly focused on behavioral interventions, such as to-do managers \cite{wu2021dillydally}, time or project tracking tools \cite{valladares2016designing, higashi2023personalized}. These approaches align with the common view of procrastination as primarily a self-regulation failure and time management issue \cite{schouwenburg2001study, wolters2017examining}. By enhancing users' motivation or facilitating their time management, these tools have been shown to improve productivity. However, procrastination is deeply rooted in psychological factors, such as irrational beliefs and negative emotions like fear of failure or low self-esteem \cite{beck2000correlates}. Existing tools offer limited support for individuals to understand these underlying psychological issues. This limitation becomes particularly problematic for those grappling with emotionally-driven procrastination. When time management tools alone prove ineffective, they may exacerbate feelings of frustration and helplessness, as individuals perceive their inability to use these tools as further evidence of personal failure \cite{rozental2014understanding}.} 

\hl{In contrast to existing behavior-oriented interventions in HCI, our study demonstrated how a serious game can approach procrastination with an educational perspective, which \textit{helped players develop both deeper understanding and a healthier emotional relationship with their procrastination behaviors}. Specifically, our findings revealed that participants developed the ability to identify, analyze, and respond to their procrastination patterns in a more structured way. Moreover, this educational approach helped players reframe their perception of procrastination from a quite negative behavior to a more nuanced challenge that could be understood and managed, which also supported them alleviate the negative emotions associated with procrastination. Notably, players reported actively experimenting with new strategies to manage their procrastination in real-life contexts, suggesting that the insights gained through gameplay had impact beyond the virtual environment. }

\hl{These results align with the understanding that meaningful behavioral transformation often follows cognitive and emotional shifts \cite{beck1970cognitive}. Therefore, we argue that this educational game-based approach could be valuable for \textit{addressing psychological challenges beyond procrastination}. Behaviors closely tied to emotional regulation, such as emotional eating and impostor syndrome, may also benefit from similar targeted learning experiences that help individuals understand underlying causes and develop effective coping strategies. By fostering cognitive and emotional shifts through engaging educational experiences, individuals may have more autonomy to explore and understand their behaviors in a more compassionate and constructive manner. }

\hl{Furthermore, \textit{ProcrastiMate} also showcased the promising potential of serious games as interactive educational tools for understanding procrastination and managing the associated thoughts and emotions. Compared to traditional methods like lectures, online videos, or coaching, we reflected that \textit{ProcrastiMate}'s game-based approach offers several distinct advantages. First, \textit{role-play demonstrated its value in serious games}. In our case, allowing players to assume the role of a counselor helping others enabled them to explore key concepts freely and without emotional burden, fostering active, experiential learning over passive knowledge transfer. Second, \textit{serious games provide an entrance for customized and adaptive in-game narratives}, with \textit{ProcrastiMate} tailoring the learning experience to each player's choices and understanding, overcoming the limitations of one-size-fits-all educational methods. Finally, we believe that the game's format has the potential to \textit{democratize access to procrastination education} by removing barriers such as cost, scheduling constraints, and stigma—particularly important given that procrastination is a widespread issue. Although \textit{ProcrastiMate} was developed as a PC-based application, future iterations could explore mobile or augmented reality (AR) platforms to deliver educational content in more immediate, everyday contexts. }

\subsection{Balancing Personal Relevance with Psychological Distance for Safe and Effective Learning}

\hl{Creating experiences that facilitate relatable engagement while encouraging learning and reflection is always crucial in serious game design. To this end, prior HCI research has extensively explored how to bring players \textit{closer} to the game environment, such as simulating real-world settings \cite{gerling2014effects}, creating relatable in-game content \cite{orji2017towards}, employing immersive first-person perspectives \cite{denisova2015first} and using second-person storytelling \cite{bellini2020choice}. For example, Harrigan and Wardrip-Fruin \cite{harrigan2010second} proposed that the second-person perspectives (i.e., using ‘\textit{you}’ as the pronoun) would engage players as active actors in game, a concept that inspired Bellini et al.'s serious game design for domestic violence interventions \cite{bellini2020choice}. However, our formative study revealed that: when dealing with sensitive psychological issues like procrastination, using the pronoun ‘\textit{you}’ could inadvertently alienate players from the intended design goals.} 

We try to interpret why direct approaches might impede learning in this context through the lens of cognitive dissonance theory \cite{cooper2007cognitive,aronson1969theory}. First, when players encountered strategies in the game that differed from their personal approaches to overcoming procrastination—methods they deemed effective—they experienced psychological discomfort, often manifesting as resistance or frustration. \hl{Additionally, procrastination itself is closely tied to negative emotions, such as guilt or shame, which can make individuals more sensitive to perceived criticism or judgment.} The use of direct second-person phrasing like “\textit{you are procrastinating...}” may have exacerbated this discomfort by inadvertently activating self-defense mechanisms, leading players to feel criticized or blamed for their behavior. This reaction often results in increased resistance, reinforcing cognitive dissonance \cite{sherman2006psychology}. In turn, these defensive reactions would reduce player engagement and negatively impact their learning experience and outcomes. 

To address this, we achieved the balance between personal relevance and psychological distance with three design considerations shown in \autoref{teaser}. Specifically, our approach \textit{shifted the player's role from that of a protagonist to a helper or counselor, which helped mitigate cognitive dissonance}. This reframing allowed players to approach procrastination scenarios with greater emotional distance, reducing the sense of personal criticism. Additionally, the design of \textit{NPCs as the voice of the game's framework served as a cognitive buffer}, enabling players to engage with new ideas without feeling as though their self-perceptions were being directly challenged. This design choice may explain how \textit{ProcrastiMate} fostered a safe learning environment, where players could explore alternative strategies with a more open mindset. While maintaining this psychological distance, we \textit{transformed players' own procrastination narratives into those of other characters to preserve relatability}, ensuring the personal relevance. 

\hl{In conclusion, \textit{ProcrastiMate} showcased how to \textit{keep players engaged without triggering defensiveness}, enabling them to \textit{learn new concepts without feeling criticized}. We argue that fostering a psychologically safe environment in serious games is a valuable consideration for future research, particularly when tackling sensitive topics where players might otherwise feel judged or vulnerable}, such as mental health challenges (e.g., anxiety, depression, addiction).

\subsection{Challenges in Translating or Connecting In-game Narratives to Personal Life}
\hl{The tensions and challenges described in \autoref{finding3}, revealed the complex relationship between knowledge acquisition (\textit{learning about behavior}) and behavioral change (\textit{acting on that knowledge}) in serious games \cite{kim2021effects}. While knowing why we procrastinate helped players confront their behaviors and prompted some to take real-life actions and experiments, \textit{it remains unclear how to reinforce this awareness to achieve sustained behavioral change}. Furthermore, \textit{how to prevent justification of procrastination through raised awareness} is another unresolved question. These topics can be crucial for procrastination intervention, though they lie beyond the scope of the current research. }

\hl{We reflect that this highlights a fundamental challenge in educational game design: while games excel at making learning engaging and providing immediate feedback \cite{flemban2024assessing}, translating this learning into sustained behavioral change requires additional support mechanisms \cite{slovak2015teaching}. Despite these challenges, the real-life experimentation reported from participants offers valuable inspirations for future research. It suggests that \textit{integrating gameplay elements into users' daily routines}—possibly through tangible formats—could support the transition from awareness to sustained change. Features such as daily reminders, progress tracking, and social reinforcement may help bridge this gap. The similar ideas can also be found in recent HCI research on habit formation \cite{kim2023routineaid, sadprasid2024leveraging}.}

\hl{Moreover, the tensions between player-defined and game-defined interpretations of procrastination highlighted the importance of \textit{incorporating value judgment into serious game design}. Our findings showed that this value-driven perspective influenced how participants engaged with the game's content, potentially leading to frustration. Based on these insights, we suggest future research integrate mechanisms that allow players to express and explore their own value judgments or create dialogue between personal beliefs and theoretical perspectives. When designed appropriately, such mechanisms are promising in fostering reflection. Given that players’ judgment may evolve during gameplay, incorporating more flexible and adaptive storyline may also be considered to enhance the learning experience. We believe that these directions are promising due to the existing HCI research has demonstrated the capability of LLMs for facilitating more dynamic, personalized gaming narratives and experiences \cite{isaza2024prompt, steenstra2024engaging}.}

\subsection{Limitations and Future Work}
Finally, several limitations of our study should be acknowledged. First, our evaluation was limited to a two-week period. Future research should investigate the long-term effects of \textit{ProcrastiMate}, which would also help address some of the issues we mentioned in Discussion. Second, our design and evaluation focused on college students, which may limit the generalizability of our findings. Future studies could extend the scenarios to workplace settings and other contexts. \hl{Third, although we assessed participants' trait procrastination tendencies in our pre-interview to better understand our sample, we did not use these scores to screen participants. As a result, the distribution of procrastination tendencies in our final sample (ranging from high to low tendencies, with only 3 participants in the low tendency group, or 11\%) was imbalanced. This limited our ability to conduct detailed statistical analysis, such as exploring how the game affects individuals with varying levels of procrastination tendencies}. Finally, we did not account for participants' prior gaming experience, which may have influenced their engagement with the game.

\section{Conclusion}
In this study, we explored how the text-adventure serious game \textit{ProcrastiMate} can help college students learn about procrastination, with key design considerations centered on balancing personal relevance and psychological distance. A two-week field study with 27 participants provided valuable insights into how \textit{ProcrastiMate} facilitated the learning process, enabling players to understand and reflect on procrastination, while also revealing how design elements influenced their gaming and learning experiences. Our study contributes to the HCI community by offering a novel perspective—game-based educational approaches—for procrastination intervention, and highlighting challenges and design implications that can inform future HCI research on similar psychological issues beyond procrastination.

\section{Acknowledgments}
This work was supported by National Natural Science Foundation of China (62202335); Sino-German Cooperation 2.0 Funding Program of Tongji University (ZD2023029); National Key R\&D Program Funded Projects (2023YFC3603502). Besides, we would like to thank Yao Wang, Xian Xu, Dingdong Liu, Yuanhao Zhang, Yuying Tang, Jiawen Liu, Assem Zhunis, Qiaoyi Chen, and Shuchang Xu for their insightful comments and suggestions on our manuscript. We also extend our gratitude to Yuqing Zhang, Jiayi Chen, Jialin Yuan, Tianyue Yang, Leheng Chen, and Nianchong Qu for their valuable feedback on the initial game ideas. Additionally, we sincerely thank all participants for their time and efforts. Finally, we express our gratitude to the reviewers for their constructive feedback, which had significantly improved the quality of our paper.

\bibliographystyle{ACM-Reference-Format}
\bibliography{sample-base}

\newpage
\appendix
\renewcommand{\thesection}{\Alph{section}}

\section{40 Coping Strategy Cards Categorized by Four Causes of Procrastination }\
\label{appendix_card}
In this appendix, we present the 40 coping strategy cards designed for our game, \textit{ProcrastiMate}. The card content has been translated into English, and each card is assigned a unique number. For literature references supporting these coping strategies, please refer to our supplementary materials.

\begin{figure*}[h]
  \centering
  \includegraphics[width=\linewidth]{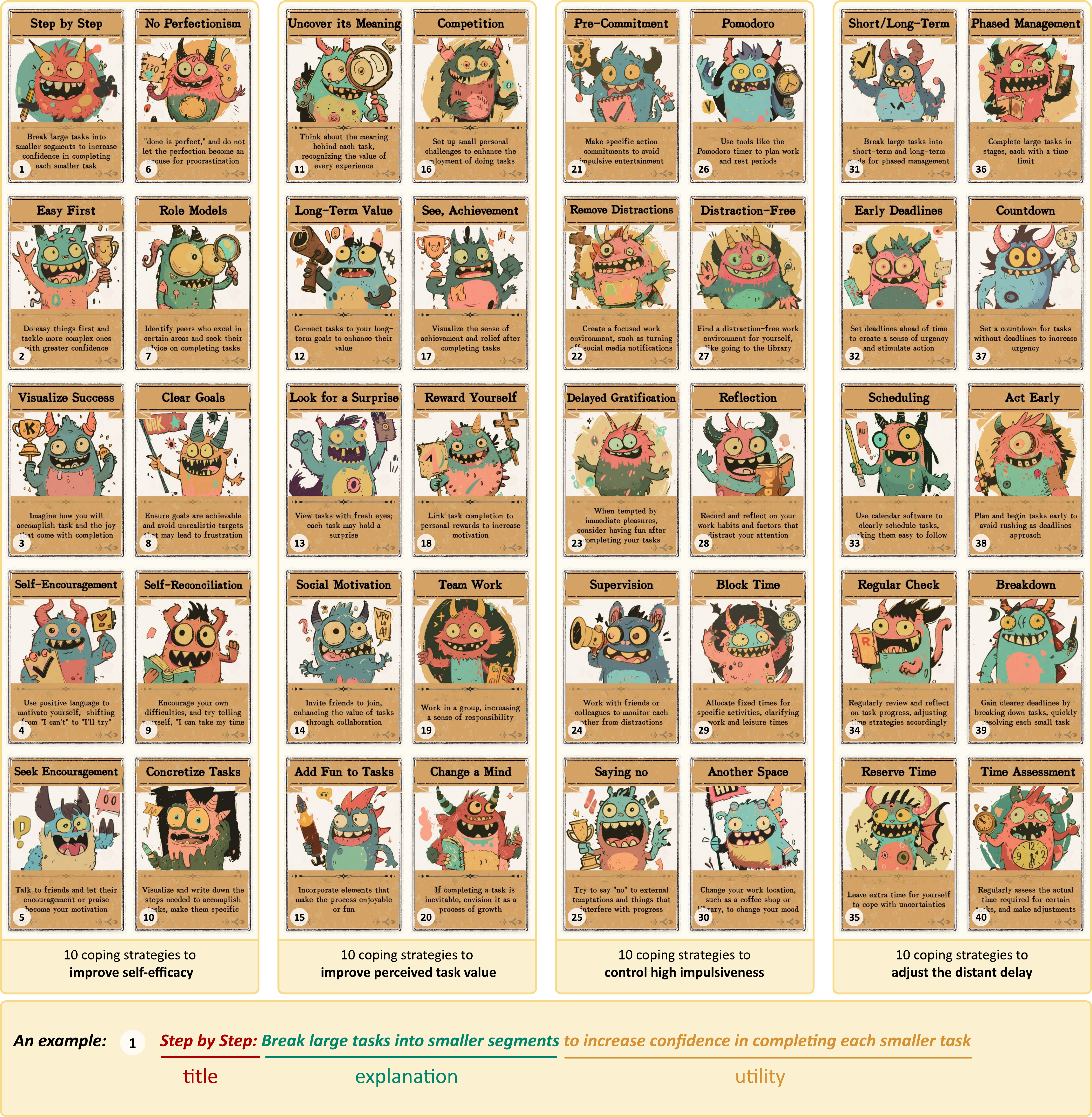}
  \caption{Upper: A deck of 40 coping strategy cards, with 10 strategies corresponding to each of the four main causes of procrastination. Lower: Taking card No.1 as an example. }
  \label{card}
\end{figure*}

\end{CJK*}
\end{document}